\documentclass[graybox]{svmult}

\usepackage{mathptmx}       
\usepackage{helvet}         
\usepackage{courier}        
\usepackage{gensymb}
\usepackage{makeidx}         
\usepackage{graphicx}        
\usepackage{multicol}        
\usepackage[bottom]{footmisc}
\usepackage{natbib}

\input{journalnames.sty}      

\makeindex             

\begin{document}

\title*{The intrinsic shape of galaxy bulges}
\author{J. M\'endez-Abreu}
\institute{J. M\'endez-Abreu \at School of Physics and Astronomy, University of St Andrews,
North Haugh, St Andrews, KY16 9SS, UK, \email{jma20@st-andrews.ac.uk}}
%
%
\maketitle

\abstract*{The  knowledge  of   the  intrinsic  three-dimensional  (3D)
  structure of  galaxy components  provides crucial  information about
  the physical  processes driving  their formation and  evolution.  In
  this paper I discuss the main  developments and results in the quest
  to  better understand  the 3D  shape of  galaxy bulges.  I start  by
  establishing the  basic geometrical description of  the problem. Our
  understanding  of the  intrinsic  shape of  elliptical galaxies  and
  galaxy discs is then presented in  a historical context, in order to
  place the role  that the 3D structure of bulges  play in the broader
  picture of  galaxy evolution.  Our current  view on the 3D  shape of
  the  Milky Way  bulge and  future prospects  in the  field are  also
  depicted.}

\abstract{The  knowledge  of   the  intrinsic  three-dimensional  (3D)
  structure of  galaxy components  provides crucial  information about
  the physical  processes driving  their formation and  evolution.  In
  this paper I discuss the main  developments and results in the quest
  to  better understand  the 3D  shape of  galaxy bulges.  I start  by
  establishing the  basic geometrical description of  the problem. Our
  understanding  of the  intrinsic  shape of  elliptical galaxies  and
  galaxy discs is then presented in  a historical context, in order to
  place the role  that the 3D structure of bulges  play in the broader
  picture of  galaxy evolution.  Our current  view on the 3D  shape of
  the  Milky Way  bulge and  future prospects  in the  field are  also
  depicted.}

\section{Introduction and overview}

Galaxies  are  three-dimensional  (3D)  structures  moving  under  the
dictates of gravity in a 3D  Universe. From our position on the Earth,
astronomers  have only  the  opportunity to  observe their  properties
projected onto a two-dimensional (2D)  plane, usually called the plane
of the  sky.  Since  we can neither  circumnavigate galaxies  nor wait
until  they spin  around,  our  knowledge of  the  intrinsic shape  of
galaxies  is still  limited, relying  on sensible,  but sometimes  not
accurate, physical and geometrical hypotheses.

Despite the obvious difficulties inherent  to measure the intrinsic 3D
shape of galaxies,  it is doubtless that it keeps  an invaluable piece
of  information  about  their   formation  and  evolution.   In  fact,
astronomers have acknowledged this  since galaxies were established to
be {\it  island universes} and  the topic has produced  an outstanding
amount of literature during the last century.

In this paper I discuss the main developments and results in the quest
to better understand the 3D shape of galaxy bulges.  Given the limited
space available in this chapter, I  have not elaborated on the concept
and definition of a bulge,  leaving this discussion to another chapter
in this volume.  In the same way, I have deliberately not included the
intrinsic shape of boxy/peanut (B/P)  structures located in the centre
of  disc  galaxies  which  some authors  associate  to  galaxy  bulges
\citep{lutticke00}.   Currently  it  is well  established  that  these
structures  are actually  part of  the bar  and intimately  related to
their  secular  evolution  \citep{combessanders81,chungbureau04}.   As
bars evolve, stars can be moved perpendicular to the disc plane due to
a  coherent bending  of  the bar  producing  its characteristic  shape
\citep{debattista04, martinezvalpuesta06}.   B/P structures  share the
same     photometric    and     kinematic    properties     of    bars
\citep{mendezabreu08b,erwindebattista13}.

On  the  other hand,  I  have  included  a  historical review  of  the
evolution  of  our knowledge  of  the  intrinsic shape  of  elliptical
galaxies.   The  properties  of   elliptical  galaxies  and  those  of
intermediate/massive galaxy  bulges have  been often considered  to be
similar \citep{wyse97}.   This is particularly true  when referring to
their  surface-brightness distributions  and shapes.   Indeed, it  has
been  common  in  the  literature  to rely  on  both  simulations  and
observations  of elliptical  galaxies to  interpret the  observational
properties of bulges \citep[e.g.,][]{kormendybender12}.

This paper  is structured  as follows.   In Section  \ref{sec:scene} I
describe the basic geometric considerations  of the problem and set up
the notation used throughout the chapter.  In Section \ref{sec:back} I
review our current knowledge on the intrinsic shape of both elliptical
and disc galaxies.  Section \ref{sec:bulges} introduces the advantages
and drawbacks  of studying galaxy  bulges with respect  to ellipticals
and  a historical  perspective  of their  3D  shape measurements.   In
Section \ref{sec:MW} I  summarize the evolution of the  concept of the
Milky   Way    bulge   and   its   intrinsic    3D   shape.    Section
\ref{sec:simulations}   addresses   the    importance   of   numerical
simulations  to understand  the physical  processes that  shape galaxy
ellipsoids.  Finally,  in Section  \ref{sec:conclusions} I  sketch out
the current view  on the intrinsic shape of bulges  and explore future
prospects.

\section{Setting up the Scene}
\label{sec:scene}

This  section briefly  summarizes the  basic notation  and geometrical
considerations to be used during this chapter.

Let ($x, y,  z$) be the Cartesian coordinates on  the reference system
of the galaxy  with the origin in the galaxy  centre, the $x-$axis and
$y-$axis  corresponding  to  the  principal  equatorial  axes  of  the
ellipsoidal  component, and  the  $z-$axis corresponding  to the  polar
axis. Therefore, if $A$, $B$, and $C$ are the intrinsic lengths of the
ellipsoid semi-axes,  the corresponding equation  of the bulge  on its
own reference system is given by

\begin{equation} 
\frac{x^2}{A^2} + \frac{y^2}{B^2} + \frac{z^2}{C^2} = 1
\label{eqn:ellipsoid} 
\end{equation} 
%
Let  $(x',y',z')$ now  be the  Cartesian coordinates  on the  observer
reference system.  It  has its origin in the galaxy  centre, the polar
$z'-$axis  is along  the line  of sight  (LOS) and  points toward  the
galaxy. $(x',y')$ represents the plane of the sky.

\begin{figure}[t]
  \includegraphics[width=\textwidth]{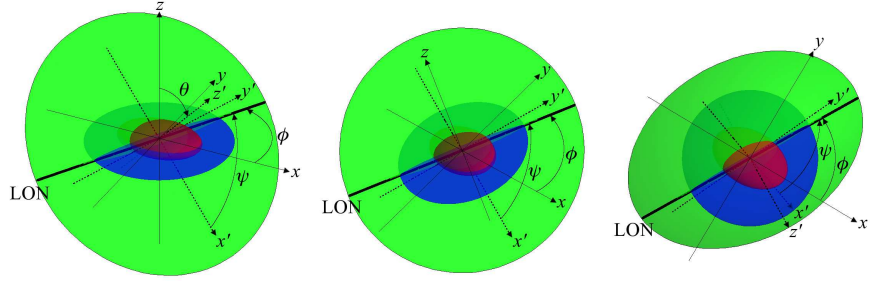}
  \caption{Schematic three-dimensional view of the ellipsoid geometry.
    The bulge ellipsoid,  the disc plane, and the sky  plane are shown
    in red, blue,  and green, respectively.  The  reference systems of
    both the ellipsoid and the observer as well as the LON are plotted
    with thin solid lines, thin dashed  lines, and a thick solid line,
    respectively.   The  bulge ellipsoid  is  shown  as seen  from  an
    arbitrary  viewing  angle (left  panel),  along  the LOS  (central
    panel),  and  along the  polar  axis  (i.e., the  $z-$axis;  right
    panel).  Extracted  from  \citep{mendezabreu10}.  Reproduced  with
    permission from Astronomy \& Astrophysics, \textcircled{c} ESO.}
  \label{fig:geometry}
\end{figure}

The equatorial  plane $(x, y)$ of  the ellipsoid and the  plane of the
sky  $(x',y')$ intersect  in the  so-called line  of nodes  (LON). The
angle between both  planes, i.e., the angle subtended  between $z$ and
$z'$  is defined  as the  inclination $\theta$  of the  ellipsoid. The
remaining two Euler angles which allow for the transformation from the
reference system of the  galaxy to that of the sky  are defined as: i)
$\phi$ is the angle subtended between  the $x-$axis and the LON in the
ellipsoid  equatorial plane,  and ii)  $\psi$ is  the angle  subtended
between the  $x'-$axis and  the LON in  the plane of  the sky.   It is
often useful to choose the $x'-$axis to be along the LON, consequently
it holds that $\psi=0$ (see Figure \ref{fig:geometry}).

It is well known that the  projection of a triaxial ellipsoid onto the
plane       of      the       sky      describes       an      ellipse
\citep{contopoulos56,stark77,binney85,franx91},   which   is   usually
written as

\begin{equation} 
\frac{x_{\rm e}^2}{a^2} + \frac{y_{\rm e}^2}{b^2} = 1
\label{eqn:ellipse} 
\end{equation} 
%
where $x_{\rm  e}$ and $y_{\rm e}$  represent the axes of  symmetry of
the projected  ellipse, $a$ and  $b$ are the  corresponding semi-major
and semi-minor axes  of the ellipse.  The observed  ellipticity of the
ellipse  can  be  easily  derived  from the  apparent  axis  ratio  as
$\epsilon=1 -  b/a$.  The  $x_{\rm e}$− axis  forms an  angle $\delta$
with the LON  (twist angle), which for convenience is  usually made to
correspond  with the  $x'$-axis.  It  is  worth noting  that both  the
apparent  axis ratio  ($q=b/a$) and  the orientation  of the  ellipses
($\delta$) depend  only, and  unambiguously, on  the direction  of the
LOS, i.e., on $\theta$, $\phi$, and $\psi$, and on the intrinsic shape
of the ellipsoid,  i.e., $A, B,$ and $C$,  see \citet{simonneau98} for
the full derivation.

Based on this  simple geometric representation, if we  assume a galaxy
is composed  of a set  of triaxial emitting ellipsoidal  shells, which
are concentric and coaxial (same  axes of symmetry) but non-homologous
(intrinsic  semi-axes vary  with the  distance to  the centre),  their
projections onto  the plane  of the sky  are concentric  ellipses, but
non-homologous and non-coaxial. Therefore,  the twisting of the galaxy
isophotes can  be explained  just as  an effect  of the  projection of
non-homologous triaxial ellipsoids \citep{williamsschwarzchild79}.

\section{Historical background on the intrinsic shape of galaxies}
\label{sec:back}

Elliptical  galaxies are  structurally  the  simplest stellar  systems
where  mathematical  techniques  can   be  applied  to  recover  their
intrinsic  3D shape.   Thus,  the  huge amount  of  literature on  the
subject  is  not surprising.   In  fact,  the continuously  increasing
availability of  better measurements  of the  apparent axis  ratios of
elliptical galaxies  have motivated great  debate over the  years.  On
the other hand, the similarities between the photometric properties of
intermediate/massive bulges and ellipticals \citep[e.g.,][]{gadotti09}
have  usually  motivated  an  extrapolation  of  the  results  on  the
intrinsic 3D  shape of  ellipticals and  their implications  on galaxy
formation and  evolution onto  the bulges of  disc galaxies.   In this
section  I revisit  our current  knowledge on  the intrinsic  shape of
elliptical galaxies  (Sect.  \ref{sec:ellipticals}) and, for  the sake
of completeness,  of disc galaxies  (Sect. \ref{sec:discs}) to  put in
context the historical background on the intrinsic shape of bulges.

\subsection{Intrinsic shape of elliptical galaxies}
\label{sec:ellipticals}

\subsubsection{Photometric approach}

The first attempt to derive the intrinsic shape of elliptical galaxies
was done  by \citet{hubble26}. At  that time, it was  already realized
the importance  of relying  on statistical methods  to recover  the 3D
shape  of  galaxies.   In  fact,  Hubble  obtained  the  frequency  of
intrinsic  short-to-long   axis  ratio   under  the   assumption  that
elliptical galaxies  were oblate  ellipsoids with  random orientations
with respect to the LOS.

Since then, this statistical approach  based on the measurement of the
apparent axis ratio distribution (AARD) and the assumption that the 3D
intrinsic  shape is  an  ellipsoids of  revolution,  either oblate  or
prolate, has been extensively used in the literature.  For the sake of
clarity I briefly outline here the basic statistical concepts.

Let us  assume the basic  geometry proposed in  Sect.  \ref{sec:scene}
and  define both  the  intrinsic ellipticity,  $Q=B/A$, and  intrinsic
flattening, $F=C/A$,  of the ellipsoid as  the corresponding intrinsic
axis    ratios     in    the     $(x,y)$    and     $(x,z)$    planes,
respectively. Therefore, in  the case of either a  pure oblate ($Q=1$)
or pure prolate  ($Q=F$) ellipsoid the Eq.  \ref{eqn:ellipsoid} can be
described by one single parameter.  If the polar axis of the ellipsoid
forms an  angle ($\theta$) with respect  to the LOS then  the apparent
axis ratio of the projected ellipse can be written as

\begin{equation}
F^2\,\sin^2{\theta}\,+\,\cos^2{\theta}= \left\{ 
  \begin{array}{l l}
    q^2    & \quad {\rm if\,\, oblate}\\
    q^{-2} & \quad {\rm  if\,\, prolate}\\
  \end{array} \right.
\label{eqn:q}
\end{equation}

Under the  realistic assumption  of randomly  distributed orientations
and  using  Eq.  \ref{eqn:q}   where  $q=q(\theta)$, the  probability
$P(q|F)dq$ that  a galaxy  with intrinsic axis  ratio $F$  is observed
with an apparent axis ratio in the range ($q, q+dq$) is

\begin{equation}
P(q|F)dq=\frac{\sin{\theta}\,dq}{|dq/d\theta|}
\label{eqn:probqF}
\end{equation}

At this point, the AARD  $\zeta(q)$,  can be  related  to  the  intrinsic
probability distribution $\xi(F)$ by

\begin{equation}
\zeta(q) = \int_0^1 P(q|F)\,\xi(F)\,dF
\label{eqn:probq}
\end{equation}

The relation  between the  known (observed)  frequency of  galaxies of
apparent axis  ratio $\zeta(q)$ to  the unknown frequency  $\xi(F)$ of
galaxies with intrinsic axis ratio $F$ can be written such as

\begin{equation}
\zeta(q)= \left\{ 
  \begin{array}{l l}
   q\,\displaystyle\int_0^q \frac{\xi(F)dF}{\sqrt{(1-F^2)(q^2-F^2)}}    & \quad {\rm if\,\, oblate} \\
    q^{-2}\,\displaystyle\int_0^q \frac{\xi(F)F^2\,dF}{\sqrt{(1-F^2)(q^2-F^2)}} & \quad {\rm  if\,\, prolate}\\
  \end{array} \right.
\label{eqn:q2}
\end{equation}

\begin{figure}[t]
\begin{center}
  \includegraphics[width=0.8\textwidth]{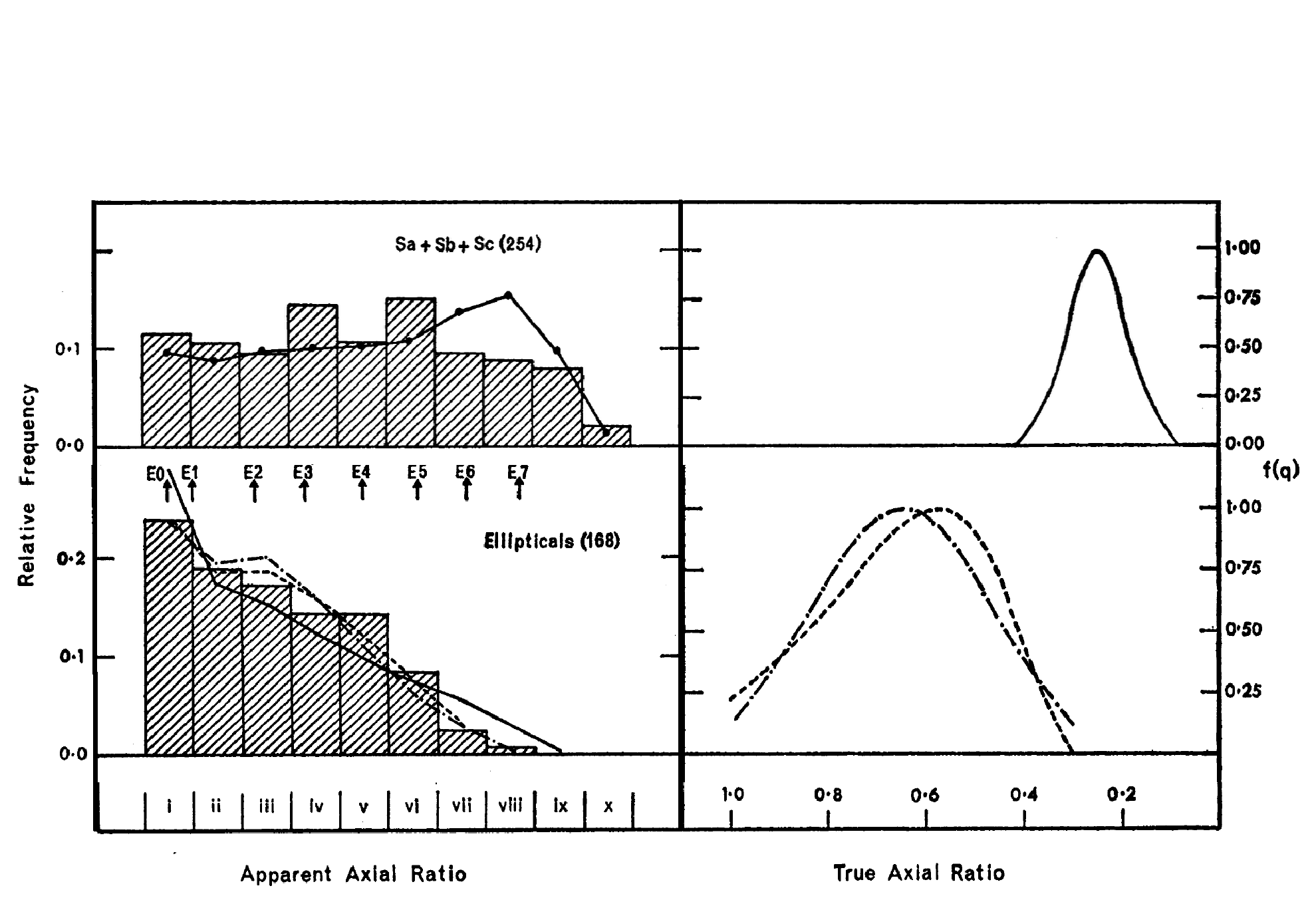}\\
  \includegraphics[width=0.75\textwidth]{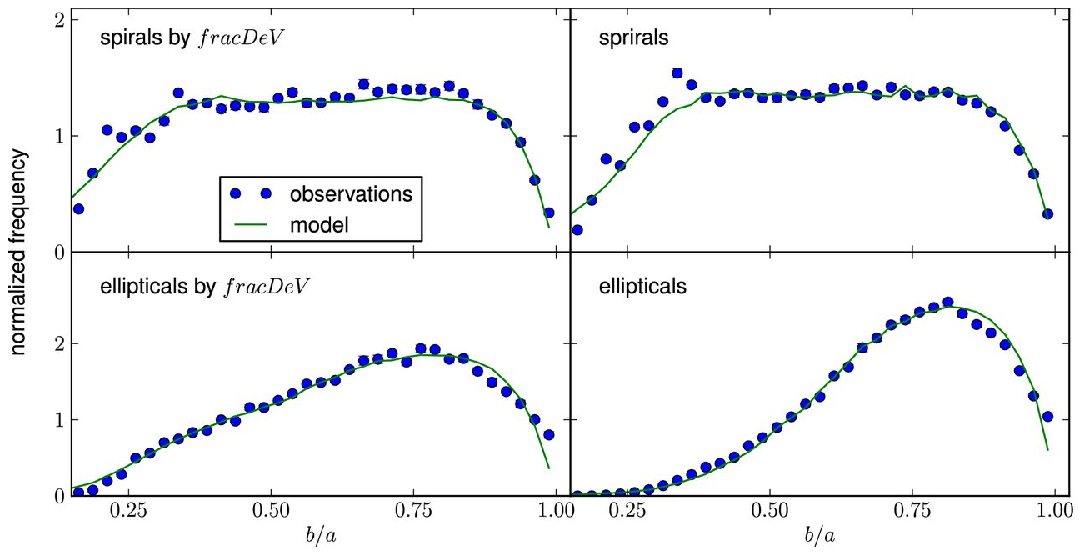}
  \caption{Composite  figure  showing  the  evolution  of  the  galaxy
    samples  used  in  the  derivation   of  the  intrinsic  shape  of
    ellipticals and  discs. Upper panels:  histograms of the  AARD for
    ellipticals  and  spiral  galaxies.  The  overplotted  curves  are
    predicted ratios  for various  assumptions of the  distribution of
    intrinsic  flattening.   On  the   right,  the  assumed  intrinsic
    distribution corresponding  to the curves on  the left.  Extracted
    from     \citet{sandage70}.     Reproduced     with    permission,
    \textcircled{c}  AAS.   Bottom panels:  best  fit  models to  AARD
    compared to the observations.  Top: spirals.  Bottom: ellipticals.
    Left: galaxies  selected only by fracDeV,  see \citet{abazajian05}
    for definition.  Right: galaxies selected by Galaxy Zoo morphology
    and              fracDeV.              Extracted              from
    \citet{rodriguezpadilla13}.  Reproduced  by permission  of  Oxford
    University Press.}
  \label{fig:ellipticals}
\end{center}
\end{figure}

Based  on  this  approach  and using  the  hypothesis  of  oblateness,
\citet{sandage70}  derived the  intrinsic  distribution of  flattening
$\xi(F)$ for  different Hubble types  ranging from ellipticals  to Sc.
They found  that the observed  axis ratios of 168  elliptical galaxies
present   in  the   Reference   Catalog  of   Bright  Galaxies   (RC1)
\citep{devaucouleurs64} were  well reproduced  using a  skewed binomial
distribution of oblate ellipsoids given by

\begin{equation}
\xi(F)\propto\left(1+\frac{F-F_{0}}{\beta}\right)^{\alpha}\,exp\left[-\alpha\,(F-F_{0})\right]
\label{eqn:probF}
\end{equation}
%
with   main    parameters   $F_0=0.58$   and    $\beta=0.31$   (Figure
\ref{fig:ellipticals}, left panels).

\citet{binney78}  used the  same  sample but  introducing the  prolate
approach.  Adopting  the same  functional form  for $\xi(F)$  he found
values  of  $F_0=0.40$  and  $\beta=0.71$.   However,  even  if  using
arbitrary analytical  representations of  $\xi(F)$ can  turn out  in a
good  fit of  the  AARD, in  principle  they do  not  have a  physical
motivation.  This approximation  was improved by \citet{noerdlinger79}
by  solving  Eq.   \ref{eqn:q2}  using  the  non-parametric  inversion
technique proposed by \citet{lucy74}.  His  results show how under the
hypothesis    of    oblateness    the   $\xi(F)$    distribution    of
\citet{sandage70}  was correct,  but he  also noticed  that a  prolate
distribution  peaking  at  around  $F\sim0.7$  would  produce  a  good
representation of the data as well.

At the same  time, some kinematic findings led to  the suggestion that
the structure  of elliptical  galaxies can  be represented  by neither
oblate nor prolate  ellipsoids of revolution.  In fact,  the low ratio
between  rotational velocity  and  velocity dispersion  found in  flat
systems \citep{bertolacapaccioli75, illingworth77,  peterson78} or the
rotation measured  along the  minor axis  of some  elliptical galaxies
\citep{schechtergunn79} were interpreted as  resulting from a triaxial
structure. From  the photometric  point of view,  the twisting  of the
inner isophotes of elliptical galaxies  was known since the early work
of  \citet{evans51}  and it  was  latter  confirmed in  several  works
\citep{liller60, carter78, bertolagalletta79}.

As      a       consequence,      \citet{benacchiogalletta80}      and
\citet{binneydevaucouleurs81}   showed   that   the  AARD   could   be
satisfactorily  accounted  for also  in  terms  of a  distribution  of
triaxial  ellipsoids.   Nevertheless,   these  works  still  presented
significant differences in the  predicted number of spherical galaxies
mainly due to  the differences in the original  samples.  Other groups
reached similar conclusions analysing  higher quality data coming from
new CCD detectors \citep{fasanovio91}.

A new  step forward  in the  methodology to  recover the  intrinsic 3D
shape of  galaxies was done  by \citet{fallfrenk83}.  They  showed how
the  inversion  of  the  integral equations  for  oblate  and  prolate
ellipsoids   (Eq.  \ref{eqn:q2})   can  be   performed  analytically,
resulting in

\begin{equation}
\xi(F)= \frac{2}{\pi}\,\sqrt{1-F^2}\left\{ 
  \begin{array}{l l}
    \frac{1}{F}\,\int_0^F \frac{qdq}{\sqrt{F^2 - q^2}}\,\frac{d\zeta}{dq}    & \quad {\rm if\,\, oblate}\\
    \frac{1}{F^3}\,\int_0^F \frac{qdq}{\sqrt{F^2 - q^2}}\,\frac{d(q^3\zeta)}{dq} & \quad {\rm  if\,\, prolate}\\
  \end{array} \right.
\label{eqn:inversion}
\end{equation}

Using this analytical inversion and  the largest sample of galaxies to
that date (2135 elliptical galaxies), \citet{lambas92} demonstrated how
neither oblate nor prolate models could adequately reproduce the data.
Contrarily, triaxial  ellipsoids with  intrinsic axis  ratios selected
from 1D Gaussians provided an adequate  fit to the data.  They found a
best fit with $Q=0.95$ and $F=0.55$.
A similar approach was used by  \citet{ryden92} on a smaller sample of
171  elliptical  galaxies.  She  used  a  2D Gaussian  combining  both
intrinsic  axis  ratios obtaining  $Q=0.98$  and  $F=0.69$.  The  same
sample  was  later  analysed   by  \citet{tremblaymerritt95}  using  a
non-parametric  technique  to  test  the  triaxial  hypothesis.   They
confirmed previous results that  discarded a distribution of intrinsic
shapes compatible with axisymmetric ellipsoids thus favouring triaxial
distributions.  Similar conclusions were reached by \citet{ryden96} on
a larger sample using the same non-parametric approach.

During these years it became  increasingly clear that the distribution
of intrinsic flattenings of elliptical galaxies was broad and possibly
bimodal   \citep{fasanovio91,ryden92,tremblaymerritt95,ryden96}.    In
fact, combining the galaxy sample  described in \citet{ryden92} with a
new    sample   of    brightest   cluster    galaxies   (BCGs)    from
\citet{lauerpostman94}, \citet{tremblaymerritt96} found  that the AARD
of galaxies  brighter than $M_B\simeq-20$  was different from  that of
the less luminous  ones.  This reflected a difference in  the shape of
low-luminosity  and high-luminosity  ellipticals: fainter  ellipticals
are moderately  flattened and  oblate, while brighter  ellipticals are
rounder and triaxial.  Recently,  \citet{fasano10} also found that even
if both normal  ellipticals and BCGs are triaxial, the  latter tend to
have a more  prolate shape, and the tendency to  prolateness is mainly
driven by the central dominant (cD) galaxies present in their sample.

The  next  qualitative leap  in  studies  of  the intrinsic  shape  of
elliptical galaxies happened with the  advent of the Sloan Digital Sky
Survey  (SDSS). With  respect to  previous statistical  analyses, SDSS
improved not  only the  number of  galaxies under  study (an  order of
magnitude  larger)  but  also  the  quality  and  homogeneity  of  the
photometry. All these improvements allowed  to study the dependence of
the  intrinsic  shape  with  other   galaxy  properties  such  as  the
luminosity, colour,  physical size, and environment.   Using data from
the  SDSS-DR3  \citep{abazajian05} \citet{vincentryden05}  found  that
bright galaxies ($M_{r}\leq-21.84$) with a de Vaucouleurs profile have
an  AARD  consistent  with  a   triaxiality  parameter  in  the  range
$0.4<T<0.8$,   where   $T=(1-Q^2)/(1-F^2)$,    and   mean   flattening
$0.66<F<0.69$.  The faintest de Vaucouleurs galaxies are best fit with
prolate ellipsoids  ($T=1$) with mean flattening  $F=0.51$.  Using the
SDSS-DR5  \citep{adelmanmccarthy07},  \citet{kimmyi07}  were  able  to
reproduce  the AARD  by using  a combination  of oblate,  prolate, and
triaxial   galaxy   populations.    Following  the   early   work   of
\citet{tremblaymerritt96},  they  assumed  each  population  having  a
Gaussian distribution of their intrinsic axis ratios.  The best fit to
the  AARD   was  found   using  a  fraction   of  O:P:T=0.29:0.26:0.45
(Oblate:Prolate:Triaxial) with a best  triaxial distribution with axis
ratios $Q=0.92$ and $F=0.78$.   In 2008, \citet{padillastrauss08} used
the SDSS-DR6  \citep{adelmanmccarthy08} to derive the  intrinsic shape
of ellipticals  with the main  improvement of taking into  account the
effects of  dust extinction.  They  found that the AARD  of elliptical
galaxies  shows   no  dependence  on  colour,   suggesting  that  dust
extinction is not  important for this sample.  The  full population of
elliptical galaxies was well  characterized by a Gaussian distribution
in  the equatorial  ellipticity  with mean  $Q=0.89$  and a  lognormal
distribution of  the flattening with mean  $F=0.43$, which corresponds
to     slightly     oblate     ellipsoids    in     agreement     with
\citet{vincentryden05}.  In a recent paper, \citet{rodriguezpadilla13}
have  used   the  SDSS-DR8  \citep{aihara11}  and   the  morphological
information from Galaxy Zoo  \citep{lintott11} finding that elliptical
galaxies have a mean  value of $F=0.58$ (Figure \ref{fig:ellipticals},
right panels).  They concluded that the  increase in $F$ is mainly due
to the removal of the spiral galaxy contamination thanks to the Galaxy
Zoo morphologies.  A  historical summary in tabular form  of all these
measurements is shown in Table \ref{tab:historical}.

Owing to the  ill-posed problem of deriving the 3D  intrinsic shape of
elliptical  galaxies, its  historical perspective  is mainly  weighted
toward statistical methods.  As previously showed in this section, the
inventiveness of astronomers, the  development of statistical methods,
and  the  advent of  large  surveys  have significantly  improved  our
knowledge  of  the  intrinsic  shape of  elliptical  galaxies.   Other
methods based  on the  photometric study  of individual  galaxies have
also been  developed but to a  smaller extent.  One of  the pioneering
works to derive the intrinsic  shape of an individual elliptical using
its   observed   ellipticity  and   isophotal   twist   was  done   by
\citet{williams81}.   They  modelled  the elliptical  galaxy  NGC~0523
assuming a given  intrinsic density distribution and  finding that the
preferred models were prolate in the external regions but increasingly
mixed (oblate and prolate) towards  the centre.  This idea was further
developed by  other authors using  more complex models of  the density
distribution   \citep{fasano95,    thakurchakraborty01}.    In   2008,
\citet{chakraborty08} estimated  the shapes of 10  elliptical galaxies
with apparent  ellipticities $\epsilon\leq  0.3$, finding  that radial
differences in the triaxiality parameter can be tightly constrained to
values  $0.29<\Delta  T<0.54$.   \citet{chakraborty11}  extended  this
analysis to 3 very flat galaxies with ellipticity $\epsilon\sim0.3$ or
more.  They found values of the intrinsic flattening of these galaxies
around $F\sim0.5$.

\subsubsection{Kinematic approach}

Determining the distribution  of the 3D intrinsic  shape of elliptical
galaxies  is  also possible  by  combining  photometric and  kinematic
information.   In  a  first   attempt,  \citet{binney85}  used  simple
kinematical models to understand the  ratio of rotational motion along
both the major and minor isophotal axes of the galaxy.  Using a sample
of  10 ellipticals  he found  that elliptical  galaxies were  not well
represented by axisymmetric oblate or prolate models.  \citet{franx91}
revisited  this approach  by using  a larger  sample of  38 elliptical
galaxies  and studying  the  probability  distribution of  photometric
ellipticities  and  kinematics  misalignments.   In  particular,  they
explored  the  possibility that  the  angular  momentum could  not  be
aligned  with  the polar  axis  of  the galaxy  but  it  may have  any
orientation within  the plane containing  the short and the  long axis
($x,z$). They found that a variety of models was able to reproduce the
observations.    Models  with   all  galaxies   being  triaxial   with
well-aligned angular momentum were  indistinguishable from models with
all galaxies being oblate with nonaligned angular momentum.

\begin{figure}[t]
  \includegraphics[width=\textwidth]{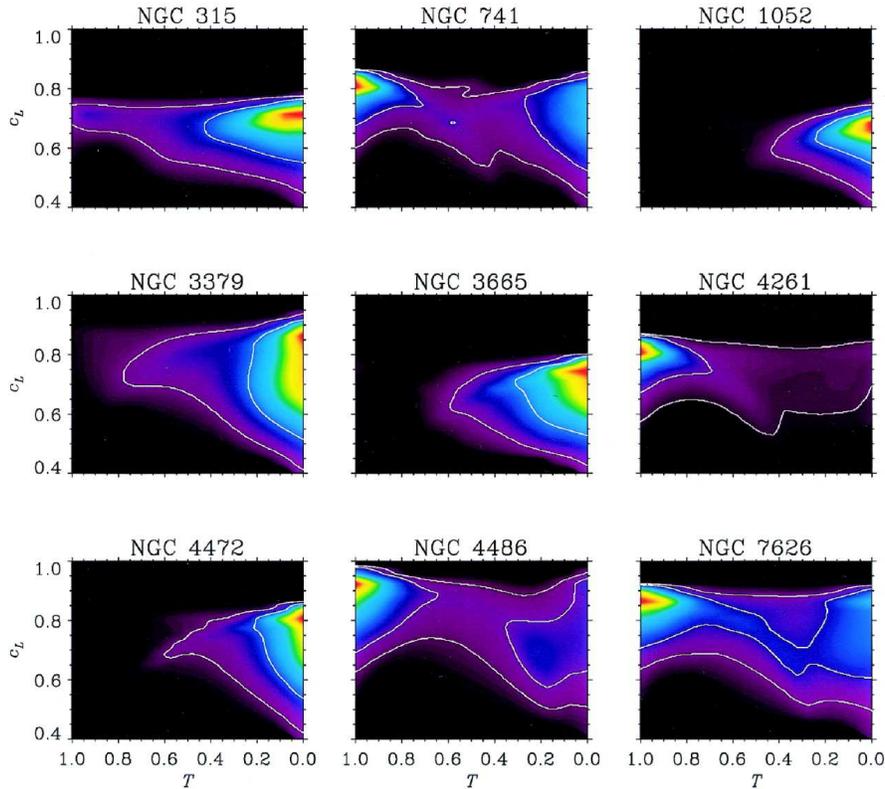}
  \caption{Posterior probability  densities in the plane  of intrinsic
    triaxiality, T, and flattening, $c_{L}$ ($F$ in this chapter), for
    each  of  the nine  galaxies  that  show significant  rotation  in
    \citet{bakstatler00}.  Contours indicate the 68\% and 95\% highest
    posterior density regions.  In  each panel, round prolate galaxies
    are at  the top left,  flattened oblate galaxies at  bottom right,
    and  objects in  between  are triaxial.   Most  galaxies are  well
    represented by  oblate models  but prolate  and triaxial  are also
    allowed in  many galaxies,  e.g., NGC~741, NGC~4486,  or NGC~7626.
    Extracted from  \citet{bakstatler00}. Reproduced  with permission,
    \textcircled{c} AAS.}
  \label{fig:ellipticals_kin}
\end{figure}

A different standpoint to statistical studies implies an investigation
into  the  intrinsic  shape  of  elliptical  galaxies  using  detailed
individual   dynamical    modelling   of   the    galaxy   kinematics.
\citet{tenjes93}  modelled  the   photometric  and  stellar  kinematic
measurements of three elliptical galaxies adopting a specific form for
the  intrinsic  density and  streaming  motions.   They found  tightly
constrained  geometries   with  $0.7<Q<0.8$  and   $0.4<F<0.6$.   This
methodology  was further  improved in  a series  of papers  by Statler
\citep{statler94a,statler94b,statler94c}. He showed how using not only
their apparent shapes  and velocity field misalignments,  but also the
velocity field asymmetry, it is  possible to place tighter constraints
on  the   intrinsic  shape   of  ellipticals.   Using   this  approach
\citet{bakstatler00}  derived the  intrinsic  shape  of 13  elliptical
galaxies  finding  that  although  photometric  studies  give  similar
results for  the flattening, none is  able to put real  constraints on
triaxiality even  when large samples are  studied, hence demonstrating
the  need   to  include   kinematic  data   in  the   models.   Figure
\ref{fig:ellipticals_kin}   show  the   probability  distribution   of
intrinsic axis ratio for 9 galaxies with significant rotation in their
sample. It is clear that most of the galaxies can be well described by
nearly oblate models but some  of them present significant triaxiality
or even  prolateness.  \citet{vandenboschvandeven09}  investigated how
well the  intrinsic shape of  elliptical galaxies can be  recovered by
fitting realistic  triaxial dynamical models to  simulated photometric
and  kinematic   observations.   They  found  that   for  axisymmetric
galaxies, the models are able to exclude triaxiality but the intrinsic
flattening is nearly  unconstrained.  On the other hand,  the shape of
triaxial  galaxies  can  be   accurately  determined  when  additional
photometric  and  kinematic  complexity,   such  as  the  presence  of
isophotal twist or a kinematically decoupled core is observed.

Recently,  \citet{weijmans14}  studied  the  intrinsic  shape  of  the
early-type    galaxies    described     in    the    ATLAS3D    survey
\citep{cappellari11}. Using a purely photometric approach and assuming
axisymmetry,  they found  that the  fast rotator  population was  much
flatter  than the  slow  rotator population,  as  expected from  their
dynamical  status.   Moreover,  when  the  kinematic  misalignment  is
included as a constraint in  the analysis, they demonstrated that fast
rotators are still better represented to oblate ellipsoids.

\begin{table}[t]
\caption{Historical summary of the intrinsic shapes of elliptical galaxies.}
\label{tab:historical}
\begin{tabular}{p{1cm}p{1.5cm}p{3cm}p{2cm}p{2cm}p{1.5cm}}
\hline\noalign{\smallskip}
   Year     &   N. Galaxies &  Hypothesis              &     $Q$       &    $F$         & Reference \\
    (1)     &       (2)     &      (3)                 &     (4)       &    (5)         &    (6)    \\
 \hline
1970        &     168       & Oblate                   & 1             & 0.58           & [1]\\
1978        &     168       & Prolate                  & 0.4           & 0.4            & [2]\\        
1979        &     168       & Oblate/Prolate           & 1/0.7         & 0.55/0.7       & [3]\\
1980        &     348       & Triaxial                 & 0.81          & 0.62           & [4]\\
1981        &     196       & Oblate/Prolate/Triaxial  & 1/0.62/0.79   & 0.62/0.62/0.57 & [5]\\ 
1992        &     2135      & Triaxial                 & 0.95          & 0.55           & [6]\\
1992        &     171       & Triaxial                 & 0.98          & 0.69           & [7]\\
2005        &     26994     & Triaxial                 & 0.66-0.85     & 0.66-0.69      & [8]\\
2007        &     3922      & Oblate/Prolate/Triaxial  & 1/0.72/0.92   & 0.44/0.72/0.78 & [9]\\
2008        &     303390    & Triaxial                 & 0.89          & 0.38           & [10]\\
2013        &     112100    & Triaxial                 & 0.88          & 0.58           & [11]\\
\noalign{\smallskip}\hline\noalign{\smallskip}
\end{tabular}
Notes. (1) Year of publication of  the paper. (2) Number of elliptical
galaxies in each sample.  (3)  Hypothesis used to derive the intrinsic
shape  of   the  ellipticals.   (4)   Mean  value  of   the  intrinsic
ellipticity. (5) Mean value of the intrinsic flattening. (6) Reference
of the corresponding paper: [1] \cite{sandage70}, [2] \cite{binney78},
[3]   \cite{noerdlinger79},    [4]   \cite{benacchiogalletta80},   [5]
\cite{binneydevaucouleurs81}, [6] \cite{lambas92}, [7] \cite{ryden92},
     [8]    \cite{vincentryden05},     [9]    \cite{kimmyi07},    [10]
     \cite{padillastrauss08}, [11] \cite{rodriguezpadilla13}.
\end{table}

\subsection{Intrinsic shape of disc galaxies}
\label{sec:discs}

In this  section I briefly  summarise our current  understanding about
the intrinsic  3D shape of  discs. Bulges  are embedded into  the disc
light and  axisymmetry is  usually a requirement  to derive  the bulge
intrinsic shape. However, although the  discs of lenticular and spiral
galaxies are often considered to be infinitesimally thin and perfectly
circular, their  intrinsic shape  is better approximated  by flattened
triaxial ellipsoids.

The   disc  flattening,   defined  analogously   as  for   ellipticals
(Sect. \ref{sec:ellipticals}), can be directly determined from edge-on
galaxies.   It depends  both  on  the wavelength  at  which discs  are
observed  and on  galaxy morphological  type.  Indeed,  galactic discs
become  thicker  at  longer  wavelengths  \citep{dalcantonbernstein02,
  mitronova04}  and   late-type  spirals   have  thicker   discs  than
early-type spirals \citep{bottinelli83, guthrie92}.

Determining  the distribution  of  both the  intrinsic flattening  and
ellipticity of discs is possible by a statistical analysis of the AARD
of randomly oriented spiral galaxies.
Similarly  for  elliptical  galaxies, \citet{sandage70}  analysed  the
spiral  galaxies listed  in the  RC1.  They  concluded that  discs are
circular with a mean flattening of $\langle F \rangle = 0.25$.
However,  the lack  of  nearly circular  spiral galaxies  ($q\simeq1$)
rules out  that discs  have a  perfectly axisymmetric  shape.  Indeed,
\citet{binggeli80},          \citet{benacchiogalletta80},          and
\citet{binneydevaucouleurs81}  have  shown  that  discs  are  slightly
elliptical with  a mean intrinsic  ellipticity $\langle 1-Q  \rangle =
0.1$.
These early findings were based on the analysis of photographic plates
of a few hundreds of galaxies.  They were later confirmed by measuring
ellipticities  of  several thousands  of  objects  in CCD  images  and
digital   scans   of   plates    obtained   in   wide-field   surveys.
\citet{lambas92} found that pure oblate models failed to reproduce the
AARD of spiral galaxies, whereas  nearly oblate models with $F\sim0.2$
and $Q\sim0.9$  produce a  good fit  with values  similar to  those of
\citet{sandage70}. These  values were confirmed later  on by different
authors \citep{fasano93, alamryden02, ryden04}.
Like  the  flattening,  the   intrinsic  ellipticity  depends  on  the
morphological type  and wavelength.   The discs of  early-type spirals
are more elliptical  than those of late-type spirals  and their median
ellipticity  increases   with  observed   wavelength  \citep{ryden06}.
Furthermore, luminous spiral galaxies tend to have thicker and rounder
discs than low-luminosity  spiral galaxies \cite{padillastrauss08}. In
\citet{sanchezjanssen10}  they studied  the  role of  stellar mass  in
shaping the thickness  of galaxy discs. They found  that the intrinsic
thickness distribution of discs has a characteristic {\it U-shape} and
identify a  limiting mass $M_{\star} \approx  2\times 10^{9}M_{\odot}$
below   which   low-mass   galaxies   start   to   be   systematically
thicker.  Recently,  \citet{rodriguezpadilla13}  analyse a  sample  of
92923 spiral  galaxies extracted  from the  SDSS-DR8, and  taking into
account  the  effects  of  dust   in  their  analysis,  they  found  a
distribution  of   flattening  with  mean  $F=0.27$   and  ellipticity
$Q=0.22$, i.e., disc  are less round than in  previous studies (Figure
\ref{fig:ellipticals}, right panels).

Despite the large effort made to  understand the intrinsic 3D shape of
galaxy  discs,  it  is  still  unclear  whether  the  inferred  slight
triaxiality could  be due  to the presence  of substructure  in galaxy
discs or if it really reflects truly triaxial potential in spirals.

\section{The intrinsic shape of extragalactic bulges.}
\label{sec:bulges}

The  study of  the intrinsic  shape of  bulges  presents similarities,
advantages, and drawbacks with respect to that of elliptical galaxies.
Bulges are ellipsoidal systems located in the centre of disc galaxies,
thus, the  main drawback with  respect to elliptical galaxies  is that
their  analysis requires  the isolation  of their  light distributions
from other structural  galaxy components. However, it  is worth noting
that a similar problem is faced in elliptical galaxies when defining a
characteristic radius to  measure the global axis ratio  of the galaxy
\citep{fasanovio91}.  The most common approach to identifying a global
axis ratio for the bulge  is by performing a photometric decomposition
of the  galaxy surface-brightness  distribution.  In this  method, the
galaxy light is usually modelled as  the sum of the contributions from
the  different  structural  components,  i.e.,  bulge  and  disc,  and
eventually     lenses,     bars,     spiral    arms,     and     rings
\citep{prieto01,laurikainen05}.     A   number    of   two-dimensional
parametric decomposition  techniques have been developed  to this aim,
such   as:  GIM2D   \citep{simard98},  GALFIT   \citep{peng02},  BUDDA
\citep{desouza04},      GASP2D     \citep{mendezabreu08},      GALPHAT
\citep{yoon11}, or IMFIT \citep{erwin15}.
On the  other hand, the main  drawback on the study  of galaxy bulges,
i.e.,  the  presence  of  other  components such  as  the  main  disc,
represents in turn  the main advantage.  The presence  of the galactic
disc allows for accurately constraining the inclination of the galaxy.
Hence, under  the assumption  that the two  components share  the same
polar axis (i.e., the equatorial plane of the disc coincides with that
of the  bulge) it allows for  the determination of the  inclination of
the bulge.  This is  crucial to  solve one of  the main  concerns when
dealing with elliptical galaxies.

\subsection{Photometric approach}
\label{sec:bulphot}

Galaxy bulges were initially thought as axisymmetric ellipsoids placed
at  the centre  of  disc  galaxies.  The  first  piece of  photometric
evidence against this idea was given by \citet{lindblad56}.  He showed
a misalignment  between the major axes  of the disc and  bulge in M31,
realising that  this would be  impossible if  both the disc  and bulge
were oblate.  This photometric misalignment is similar to the isophote
twist observed  in elliptical  galaxies and used  as an  indication of
triaxiality  in  these  systems  \citep{williamsschwarzchild79}.   The
extensive study  undergone by \citet{kent84} showed  that the twisting
isophotes between  the central  and outer parts  of disc  galaxies are
quite  common, but  it was  not until  1986 when  \citet{zaritskylo86}
properly  studied the  deviations from  axisymmetry in  the bulges  of
spiral  galaxies.  They  found  bulge-to-disc  misalignments in  their
sample of  11 spiral galaxies  hence confirming the high  incidence of
non-axisymmetric  bulges   in  ordinary   spirals  and   placing  some
parallelisms with  elliptical galaxies.  \citet{beckman91}  also found
compelling  photometric  evidence  for  triaxiality in  the  bulge  of
NGC~4736.

The first quantitative estimation of  the intrinsic 3D shape of galaxy
bulges    using   a    statistical   approach    was   performed    by
\citet{bertola91}.  They measured the bulge AARD and the misalignments
between the  major axes of  the bulge and disc  in a sample  32 S0--Sb
galaxies. Under  the hypothesis  that discs  are circular,  they found
that  these bulges  are triaxial  with  mean axial  ratios $\langle  Q
\rangle=0.86$ and $\langle F  \rangle=0.65$.  Interestingly, they also
demonstrated that a random  projection of the probability distribution
function of the  bulges axis ratios fit sufficiently well  to the AARD
of the elliptical galaxies presented in \citet{binneydevaucouleurs81}.
The results were interpreted as both populations of objects having the
same origin.

\citet{fathipeletier03} derived the intrinsic ellipticity of bulges by
analysing the deprojected apparent axis  ratio of the galaxy isophotes
within the  bulge radius.   This work did  not assume  any geometrical
model for  the galaxy but only  that the disc be  circular. They found
$\langle Q\rangle = 0.79$ and $\langle Q\rangle = 0.71$ for the bulges
of 35 early-type and 35 late-type disc galaxies, respectively. Despite
the different methodologies, these results were in good agreement with
previous results by \citet{bertola91}.  Along  the same lines, none of
the  21 disc  galaxies with  morphological  types between  S0 and  Sab
studied by \citet{noordermeervanderhulst07} harbours a truly spherical
bulge.  They  reach this  conclusion by assuming  bulges to  be oblate
ellipsoids   and   comparing  the   isophotal   axis   ratio  in   the
bulge-dominated  region   to  that  measured  in   the  disc-dominated
region. A mean flattening $\langle  F \rangle=0.55$ was obtained which
is slightly lower than the value found by \citet{bertola91}.

The  number of  galaxy bulges  under study  increased by  an order  of
magnitude with  the work of \citet{mendezabreu08}.   They measured the
structural  parameters  of  bulges  and  discs  of  a  sample  of  148
early-to-intermediate   spiral  galaxies   using   a  2D   photometric
decomposition.  They computed the probability distribution function of
the intrinsic  ellipticity from  the bulges AARD,  disc ellipticities,
and  misalignments between  bulges  and discs  position angles.   They
suggested that about 80\% of the sample bulges are triaxial ellipsoids
with a mean axial ratio $\langle  B/A \rangle$ = 0.85, confirming that
bulges are slightly triaxial structures.

The vertical  extension of galaxy  bulges remains usually  hidden from
observations except for edge-on galaxies.  \citet{mosenkov10} obtained
a median value of the flattening $\langle F \rangle=0.63$ for a sample
of  both early-  and late-type  edge-on galaxies  using near  infrared
photometry.   These results  match  well with  the  early findings  by
\citet{bertola91}.

As well as for elliptical galaxies a number of works have attempted to
quantify  the   intrinsic  shape  of  individual   bulges  using  only
photometric  data.  The  pioneering  work of  \citet{varela96} used  a
combination of  geometrical deprojection and photometric  inversion to
work out the actual shape of  the galaxy bulge in NGC~2841. They found
that  a family  of triaxial  ellipsoids with  variable axis  ratios is
necessary to explain the photometric properties of its bulge. In 1998,
\citet{simonneau98}  derived a  set  of equations  defining the  three
intrinsic axes of  a triaxial ellipsoid as a function  of the measured
geometry of a galaxy bulge and  disc (axis ratios and position angles)
and the  unknown Euler angle  $ \phi$ (see Sect.   \ref{sec:scene} for
definition).    This    seminal   paper    promoted   the    work   of
\citet{mendezabreu10}.   They introduced  a new  method to  derive the
intrinsic shape of bulges based  upon the analytical relations between
the  observed and  intrinsic shapes  of bulges  and their  surrounding
discs.   Using  the  equations   derived  in  \citet{simonneau98}  and
introducing  physical constraints  on the  accessible viewing  angles,
they found the  following relation between the  intrinsic semi-axes of
the bulge and their observed properties

\begin{figure*}
  \includegraphics[width=0.49\textwidth]{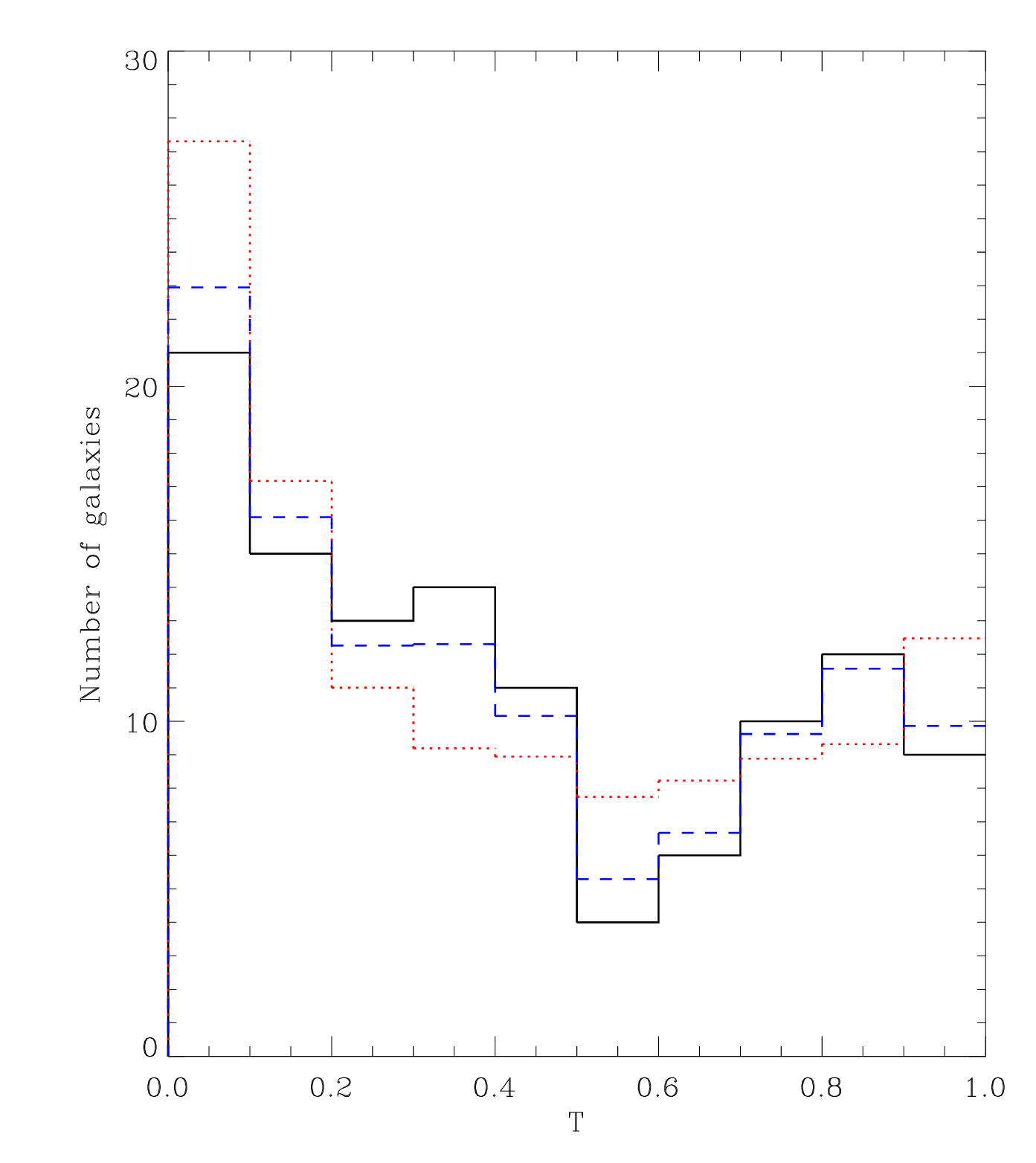}
  \includegraphics[width=0.49\textwidth]{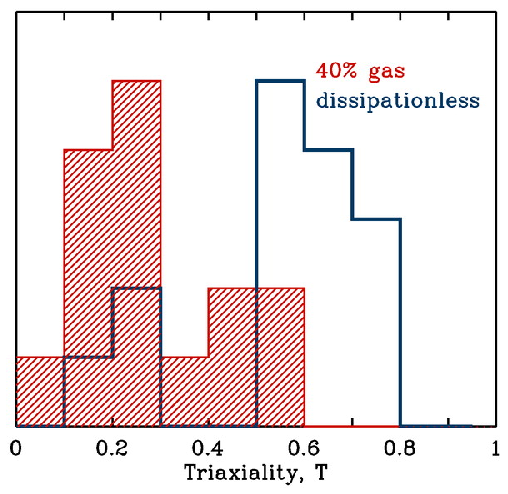}
  \caption{Composite figure  showing the similar  bimodal distribution
    of  triaxiality  parameters  from observations  (left  panel)  and
    simulations  (right  panel).   Left  panel:  distribution  of  the
    triaxiality   parameter   $T$   obtained  from   the   sample   of
    \citet{mendezabreu10} (continuous line) and for a simulated sample
    with both 30\%  and 100\% of bulges hosting a  nuclear bar (dashed
    and     dotted    lines),     respectively.     Extracted     from
    \citet{mendezabreu10}. Reproduced  with permission  from Astronomy
    \& Astrophysics,  \textcircled{c} ESO.  Right  panel: distribution
    of  both  dissipational  (hatched histogram)  and  dissipationless
    (solid   line)   mergers   remnant  triaxiality   parameter   from
    \cite{cox06}.  In both panels  oblate galaxies have $T=0$, prolate
    galaxies   have   $T=1$,   and   all   values   in   between   are
    triaxial.   Extracted   from    \citet{cox06}.   Reproduced   with
    permission, \textcircled{c} AAS.}
  \label{fig:bulgestriax}
\end{figure*}

\begin{small}
\begin{equation}
\frac{2\,\sin{\left(2\phi_C\right)}}{F_{\rm \theta}}\,F^2 = \sin{\left(2\phi_C-\phi_B\right)} \sqrt{\left(1-Q^2\right)^2 - \sin^2{\phi_B}\left(1+Q^2\right)^2}-\sin{\phi_B}\cos{\left(2\phi_C - \phi_B\right)}\left(1+Q^2\right)^2
\label{eqn:vartie}
\end{equation}
\end{small}

where $\phi_B$,  $\phi_C$ and  $F_{\rm \theta}$  are functions  of the
observed quantities  $a$, $b$,  $\delta$, and $\theta$,  see equations
12,    13,    and    43    of    \citet{mendezabreu10}.     Therefore,
Eq. \ref{eqn:vartie}  directly relates the  intrinsic 3D shape  of the
bulge  with  its  observed properties.   Unfortunately,  the  relation
between  the intrinsic  and projected  variables also  depends on  the
spatial position  of the  bulge with  respect to the  disc on  its own
reference system (i.e., on the $\phi$ angle) and therefore, as well as
for ellipticals,  a deterministic  solution of  the problem  cannot be
given.     However,    the    statistical   analysis    provided    in
\citet{mendezabreu10} allows us to obtain the probability distribution
function  of both  semi-axis ratios,  $Q$  and $F$,  for every  single
bulge, thus imposing tight constraints  on its actual shape.  Applying
this   technique    to   the    sample   of   bulges    presented   in
\citet{mendezabreu08}  they  found  a   bimodal  distribution  of  the
triaxiality parameter (Figure  \ref{fig:bulgestriax}, left panel).  In
particular, bulges  with S\'ersic  index $n \leq  2$ exhibit  a larger
fraction  of oblate  axisymmetric (or  nearly axisymmetric)  bulges, a
smaller fraction  of triaxial  bulges, and fewer  prolate axisymmetric
(or nearly axisymmetric)  bulges with respect to bulges with  $n > 2$.
Despite no  correlations being  found between  the intrinsic  shape of
bulges  and other  properties  such as  bulge  luminosity or  velocity
dispersion,   the  differences   with  the   bulge  surface-brightness
distribution hint towards the  presence of different bulge populations
as suggested by \citet{kormendykennicutt04}.

\subsection{Evidences of triaxiality from kinematic measurements}
\label{sec:bulkin}

Early kinematic  studies of  galaxy bulges were  shown to  rotate more
rapidly  than elliptical  galaxies \citep{kormendyillingworth82}.   In
fact, the  kinematic properties of  many bulges are well  described by
dynamical models of oblate ellipsoids  which are flattened by rotation
with    little    or   no    anisotropy    \citep{daviesillingworth83,
  jarvisfreeman85,  fillmore86,  corsini99,  pignatelli01}.   However,
there are  also kinematic evidences  supporting a triaxial shape  in a
non-negligible  fraction of  these  bulges. In  1989, two  independent
works  of \citet{bertola89}  and  \citet{gerhard89}  reached the  same
conclusion about the triaxial bulge of the Sa galaxy NGC~4845. Using a
combination of  photometric and  kinematic measurements  they restrict
the intrinsic axis  ratio of its bulge to $Q=0.74$  and $F=0.6$. Their
works  were   mainly  supported   by  the  presence   of  non-circular
gas-motions in the galaxy centre. In a non-axisymmetric potential, the
shape of the rotation curve will depend on the position of the LOS and
the major  axis of  the non-axisymmetric  component.  A  slowly rising
rotation curve  or one in which  a bump of extreme  velocities is seen
near     the     centre     are     indications     of     triaxiality
\citep{gerhard89}. Based on these  considerations, and building on the
early  work of  \citet{lindblad56}, \citet{berman01}  demonstrated the
presence of a triaxial bulge in  the Andromeda galaxy (M31) by using a
hydrodynamical  simulation to  match  the observed  properties of  the
galaxy.   Further  evidences for  non-circular  gas  motion in  galaxy
centres    can    be     found    in    \citet{falconbarroso06}    and
\citet{pizzella08}.   Other kinematic  evidence for  the existence  of
triaxial bulges  comes from the  presence of velocity  gradients along
the galaxy minor axis.  \citet{corsini03} found minor axis rotation in
80\%  of their  early-type  spiral  sample.  In  a  series of  papers,
\citet{coccato04, coccato05} found that  60\% of the unbarred galaxies
show  a remarkable  gas velocity  gradient along  their optical  minor
axis.  This was achieved by combining their own data with that present
in the  literature (Revised  Shapley-Ames Catalog of  Bright Galaxies)
\citep{sandagetammann81}.

Despite the  importance of  adding kinematic information  to determine
the  intrinsic shape  of  the bulges,  and contrary  to  the works  on
elliptical  galaxies  \citep[e.g.,][]{statler94a},   there  is  not  a
well-established methodology to quantify  the degree of triaxiality of
bulges using the combined photometric and kinematic information, yet.

\subsection{Polar bulges}

Polar   bulges,    as   well   as   their    analogous   polar   rings
\citep{whitmore90},  are  elongated  structures perpendicular  to  the
plane of the galaxy disc.  A common signature of both the orthogonally
decoupled  bulge systems  and the  polar  ring galaxies  is that  both
contain  a  structural  component  whose angular  momentum  vector  is
roughly parallel to the major axis of the host galaxy.

Vertical elongation is  not a common feature of  bulges.  Indeed, most
bulges   can   be  assumed   to   be   flattened  by   rotation   (see
Sect. \ref{sec:bulkin}).   Furthermore, orthogonally  decoupled bulges
are usually not even {\it allowed} in most statistical works since the
condition   $A>B>C$   is   commonly   used,   see   \citet{bertola91}.
\citet{mendezabreu10} relaxed this condition  and found that only 18\%
of the observed  bulges have a probability $>50$\%  of being elongated
along the  polar axis with  no bulges reaching a  probability $>90$\%.
In fact, to date NGC 4698 \citep{bertola99}, NGC 4672 \citep{sarzi00},
and UGC  10043 \citep{matthewsdegrijs04} are the  only spiral galaxies
known to  host a prominent  bulge sticking out  from the plane  of the
disc.

The case of NGC~4698 is particularly  intriguing since it hosts also a
polar  nuclear stellar  disc aligned  with  its polar  bulge and  thus
perpendicular to the main disc.  This galaxy was recently revisited by
\citet{corsini12}  and  its  intrinsic  shape was  derived  using  the
methodology proposed by \citet{mendezabreu10}.   They found a slightly
triaxial polar bulge elongated along  the vertical direction with axis
ratios $Q  = 0.95$ and  $F$ = 1.60. This  result agrees well  with the
observed kinematics  presented in  \citet{bertola99} and with  a model
where  the nuclear  disc  is  the end  result  of  the acquisition  of
external gas by the pre-existing triaxial bulge on the principal plane
perpendicular to its shortest axis  and perpendicular to the main disc
of the galaxy.

\section{The Intrinsic Shape of the Milky Way Bulge}
\label{sec:MW}

Owing to its vicinity, the Galactic  bulge has always been targeted as
the ideal benchmark for  structure, kinematic, and stellar populations
studies of  bulges. In fact,  it can be studied  at a unique  level of
detail, in comparison to external  galaxies, thanks to the possibility
of measuring  the properties of  individual stars.  However,  our {\it
  inside  view} of  the Galaxy  generally restricts  our knowledge  to
pencil beam  areas around the Galactic  centre due to either  the high
extinction, the crowding, or  the superposition of multiple structures
along  the  LOS,   making  studies  of  the   inner  Galactic  regions
challenging.
The structure of the Galaxy has  accounted for a significant amount of
literature in the past and the topic has come back in the limelight in
recent  years. In  this section  I briefly  review the  Galactic bulge
topic focusing  on its  intrinsic shape heading  the readers  to other
chapters in this volume for more information about its stellar content
and kinematics.

In recent  decades it  has become  clear that the  Galaxy is  a barred
system  \citep{blitzspergel91,lopezcorredoira05} and  that most  likely
its central regions are dominated by  a boxy bulge created by vertical
instabilities         within          the         Galactic         bar
\citep{dwek95,martinezvalpuestagerhard11,   ness13}.   The   historical
evolution of our knowledge of  the intrinsic structure of the Galactic
bulge has been written by  a succession of progressively larger scale,
deeper sensitivity photometric and spectroscopic surveys.

The first  attempt to understand the  shape of the Galactic  bulge was
made by \citet{devaucouleurspence78}.  They  found that models ranging
from  spherical  to $F=0.6$  were  able  to  represent well  both  the
distribution of globular  clusters around the Galactic  centre and the
infrared  isophotes  observed  at  2.4$\mu$m  \citep{maihara78}.   The
flattening of the Galactic bulge was then further constrained with the
arrival  of the  Infrared Astronomical  Satellite (IRAS).   Using IRAS
data, \citet{harmongilmore88} and  \citet{whitelock91} found values of
the   intrinsic   flattening    spanning   $0.6<F<0.8$   using   $JHK$
near-infrared bands.   Similarly, \citet{kent91} found that,  at first
order, the  Galactic bulge can  be represented by an  oblate ellipsoid
with $F=0.61$ using data from the Infrared Telescope (IRT).

\begin{figure}[t]
  \includegraphics[width=\textwidth]{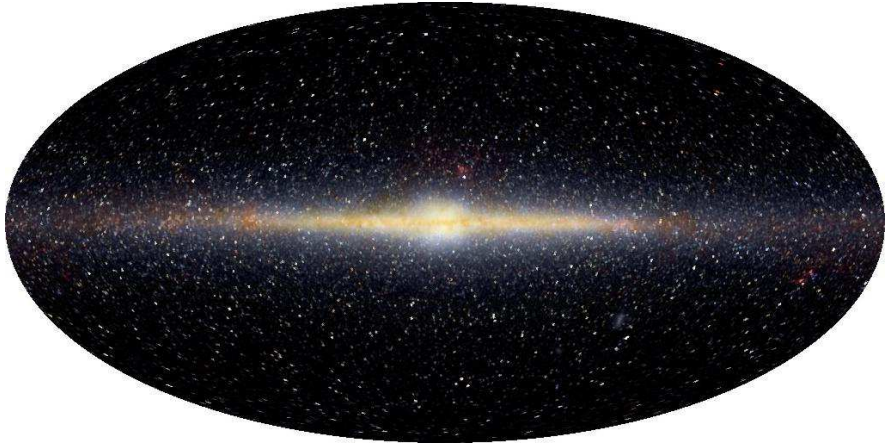}
  \caption{False-colour image of the near-infrared  sky as seen by the
    DIRBE.  Data  at  1.25,  2.2,   and  3.5  $\mu$m  wavelengths  are
    represented  respectively as  blue,  green and  red colours.   The
    image is presented in Galactic  coordinates, with the plane of the
    Milky Way  Galaxy horizontal  across the  middle and  the Galactic
    centre  at the  centre. Credits:  E. L.   Wright (UCLA),  The COBE
    Project, DIRBE, NASA.}
  \label{fig:mwcobe}
\end{figure}

The picture changed drastically with  the advent of the COBE satellite
\citep{hauser90}.  The  new striking  image of  the Milky  Way (Figure
\ref{fig:mwcobe}) provided  by the DIRBE  experiment on board  of COBE
allowed \citet{blitzspergel91}, and later  on \citet{blitz93}, to find
the  first  direct  evidence  for   a  bar  at  the  Galactic  centre.
Interestingly,  they  also found  the  presence  of a  triaxial  bulge
structurally  distinct  from the  main  bar.   The modelling  of  this
triaxial bulge was performed by different teams with different sets of
data  in the  subsequent years.   Consequently, different  axis ratios
represented as 1:Q:F were found: 1:0.33:0.22 \citep{dwek95}, 1:0.6:0.4
\citep{binney97},1:0.43:0.29       \citep{stanek97},       1:0.38:0.26
\citep{freudenreich98},     1:0.54:0.33     \citep{lopezcorredoira00},
1:(0.3--0.4):0.3          \citep{bissantzgerhard02},         1:0.5:0.4
\citep{lopezcorredoira05}.   In  general,  these  values  implied  the
Galactic  bulge  to  be  a  triaxial  structure  with  a  tendency  to
prolateness, thus  not in  agreement with the  triaxial/oblate picture
outlined in Section \ref{sec:bulges} for extragalactic bulges.
 
Although the idea of a triaxial  bulge worked well at first order, the
boxy shape  noticed earlier  by \citet{kent91} and  \citet{kent92} and
confirmed  by   \citet{dwek95}  was   not  recovered  by   a  triaxial
ellipsoid. In  the meanwhile, different  scenarios came up  to explain
these  differences   and  account  for  the   continuously  increasing
kinematic  and   stellar  populations   information.   \citet{alard01}
suggested  the  presence  of  two  different bars  in  the  Galaxy  by
analysing  data   from  the   Two  Micron   All  Sky   Survey  (2MASS)
\citep{skrutskie06}.   Another possible  scenario  was  worked out  by
\citet{babusiaux10} suggesting  a model composed by  a classical bulge
in the centre and a boxy bulge in the outer parts.

\begin{figure}[t]
  \includegraphics[width=\textwidth]{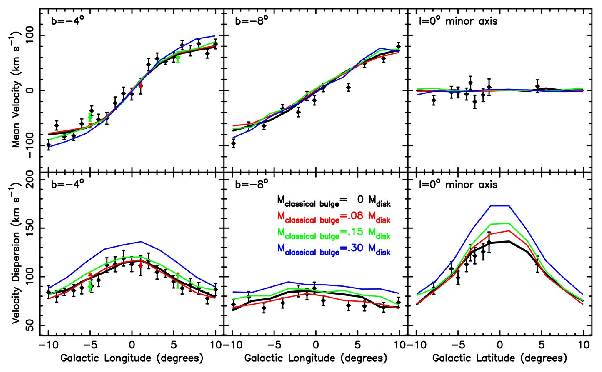}
  \caption{Best  models fits  to  the BRAVA  stellar kinematics  using
    different hypothesis  on the classical bulge  mass.  Mean velocity
    (top panels)  and velocity  dispersion (lower panels)  profiles of
    all available kinematic  observations presented in \citet{shen10}.
    The  left   two  panels  are   for  the  Galactic  latitude   b  =
    −4$^{\degree}$  strip; the  middle  two  panels are  for  the b  =
    −8$^{\degree}$;  and  the  right  two  panels  are  for  the  l  =
    0$^{\degree}$  minor axis.   The heavy  black lines  represent the
    model without a  classical bulge.  The red, green,  and blue lines
    are for  models whose classical  bulges have masses of  8\%, 15\%,
    and 30\%, respectively,  of the disk mass.   Including a classical
    bulge significantly worsens the model fits to the data, especially
    along the  minor axis.  Extracted from  \citet{shen10}. Reproduced
    with permission, \textcircled{c} AAS.}
  \label{fig:mwshen}
\end{figure}

\citet{shen10}  proposed a  simple model  yet  backed up  by the  high
quality stellar kinematics provided by the Bulge Radial Velocity Assay
(BRAVA)  \citep{rich07}.   Using  N-body  simulations  they  found  no
evidence for a classical bulge in  the Galaxy but the bulge appears to
be  only part  of the  bar and  therefore not  a separated  component.
Figure \ref{fig:mwshen} shows that the  inclusion of a classical bulge
greatly worsens the model fit to the data.  Models from \citet{shen10}
rule out  that the Milky  Way has  a significant classical  bulge with
mass $>$15\% of the disc mass.
        
Following  this line,  \citet{martinezvalpuestagerhard11} demonstrated
how  the star  counts measurements  by \citet{cabreralavers07}  agrees
with  a scenario  composed by  a single  bar and  a boxy  bulge.  More
recent measurements of star counts from the VISTA Variables in The Via
Lactea  (VVV)  \citep{gonzalez11},   metallicity  gradients  from  the
Abundances  and  Radial  velocity   Galactic  Origins  Survey  (ARGOS)
\citep{ness13},  or  stellar  kinematics  from BRAVA  have  also  been
reconciled              within               this              picture
\citep{gerhardmartinezvalpuesta12,martinezvalpuestagerhard13}.

\section{The 3D shape of bulges in numerical simulations.}
\label{sec:simulations}

The intrinsic shape of bulges  keeps important information about their
formation  history,  with  different merger,  accretion  and  assembly
scenarios  resulting in  different shapes.   Hence, the  comparison of
measured intrinsic  shapes with the output  from numerical simulations
represents  an intrinsic  way  to gain  insights  on their  formation.
However,  numerical  resolution  problems have  often  hampered  these
studies  and our  interpretation of  the shapes  of bulges  is usually
restricted to the analysis of simulated elliptical galaxies.

\citet{cox06} studied the structure  of ellipsoidal remnants formed by
either major (equal-mass) dissipationless  or dissipational mergers of
disc galaxies.  They  found a bimodal distribution  of the triaxiality
parameter  in their  remnant ellipticals  (see right  panel in  Figure
\ref{fig:bulgestriax}).  Thus,  dissipationless remnants  are triaxial
with  a tendency  to  be  more prolate  and  with  a mean  triaxiality
parameter $T=0.55$,  whereas dissipational  remnants are  triaxial and
tend  to be  much closer  to oblate  with triaxiality  $T=0.28$.  This
simulated bimodal  distribution was compared  by \citet{mendezabreu10}
to  the triaxiality  measured in  their  sample of  115 galaxy  bulges
(Figure  \ref{fig:bulgestriax}).   They   concluded  that  both  major
dissipational and dissipationless mergers  are required to explain the
variety of  shapes found for  bulges. The detailed study  presented by
\citet{cox06} is  consistent with previous studies  of dissipationless
and                        dissipational                       mergers
\citep[e.g.,][]{barnes92,hernquist92,springel00}.  However,  the study
of  \citet{gonzalezgarciabalcells05}  they  found how  the  degree  of
triaxiality of the elliptical remnants in dissipationless mergers also
depends on the morphology of  the progenitor spirals.  The presence of
central bulges on the progenitor  galaxies produce remnants which tend
to  be  more  oblate  whereas bulgeless  progenitors  lead  to  highly
triaxial   remnants  which   seems  inconsistent   with  observations.
Therefore,  the comparison  between simulations  and observations  are
still subject to the range of initial conditions explored by numerical
simulations.

\begin{figure}[t]
  \includegraphics[width=\textwidth]{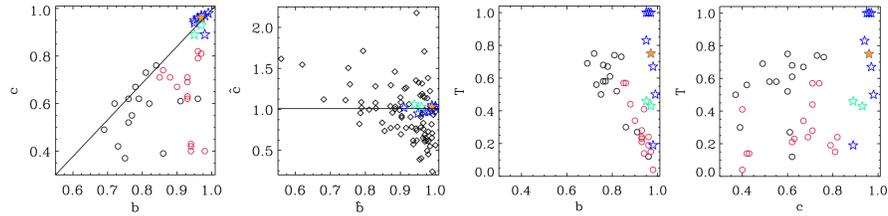}
  \caption{Intrinsic shape of bulges  and elliptical galaxies obtained
    from numerical  simulations. A comparison with  observed bulges is
    shown in the second panel. The  blue and green stars in all panels
    represent  the bulge  remnants after  suffering intermediate/minor
    mergers.   The location  of the  progenitor bulges  is shown  with
    orange stars.  The elliptical remnants  of major mergers with pure
    exponential stellar  discs (black circles) and  containing 40\% of
    gas  (red  circles)  are   also  shown.   First  panel:  intrinsic
    ellipticity  $b$  ($Q$  in  this  chapter)  versus  the  intrinsic
    flattening $c$ ($F$ in this chapter)  Second panel: as panel 1 but
    adding     the    observed     distribution    of     bulges    in
    \citet{mendezabreu10} (black diamonds).   Third and fourth panels:
    triaxiality parameter  as a function of  the intrinsic ellipticity
    and  flattening. Extracted  from \citet{tapia14}.  Reproduced with
    permission from Astronomy \& Astrophysics, \textcircled{c} ESO.}
  \label{fig:bulgestapia}
\end{figure}

On  the  other hand,  even  if  the  similarities between  bulges  and
ellipticals have prompted observers to compare the measured properties
of  bulges to  the properties  of simulated  elliptical galaxies,  the
formation path  of bulges is  likely a more complex  process involving
the    interaction   with    other   galaxy    structural   components
\citep{kormendykennicutt04,athanassoula05}.    The   recent  work   by
\citet{tapia14}  has started  to fill  the  gap on  studies about  the
intrinsic  shape of  galaxy bulges  from numerical  simulations.  They
analysed a set  of $N-$body simulations of intermediate  and minor dry
mergers onto S0s to understand  the structural and kinematic evolution
induced  by  the encounters.   In  their  experiments, the  progenitor
bulges are nearly  spherical.  The remnant bulges  remain spherical as
well ($Q \sim F >$ 0.9),  but exhibiting a wide range of triaxialities
($0.20  < T  <  1.00$), remarking  how the  definition  of this  shape
parameter  is  too  sensitive  to nearly  spherical  systems.   Figure
\ref{fig:bulgestapia} (second panel) shows how the axis ratios derived
from these simulations  (open stars) are hardly  reconcilable with the
observations  (black diamonds)  by \citet{mendezabreu10}.   Still, the
strong triaxiality  agrees with  the structure of  elliptical remnants
resulting from major-to-intermediate mergers \citep{cox06}.

\section{Concluding remarks and future prospects}
\label{sec:conclusions}

I present here a review of  our current understanding of the intrinsic
3D  shape of  galaxy  bulges. The  approach taken  in  this review  is
largely observational  and follows  the historical development  of the
field. Thus, a  journey through the past and present  of our knowledge
on the intrinsic  shape of other galaxy ellipsoids  such as elliptical
galaxies or galaxy discs was needed to put the problem in context. The
major conclusions of this review are:

\begin{itemize}
\item  The observational  data  representing the  whole population  of
  elliptical  galaxies is  consistent  with a  mixed model,  combining
  partly oblate  and partly prolate  galaxies, although a  more likely
  alternative point towards at least  some fraction of the ellipticals
  being triaxial ellipsoids.  Triaxiality is also supported by several
  photometric  and  kinematics properties,  as  well  as for  detailed
  modelling of individual galaxies.

\item The intrinsic shape of  ellipticals shows a dependence on galaxy
  luminosity.   Bright  ellipticals are  in  general  triaxial with  a
  tendency to be rounder whereas  faint ellipticals are more flattened
  with a tendency to be oblate ellipsoids.

\item Even if uncertainties due to  the lack of number statistics have
  been overcome with the advent of  recent surveys, the data can still
  be reproduced  by a wide  variety of intrinsic  shape distributions.
  Furthermore, a proper  interpretation of the data  is complicated by
  the  fact that  the AARD  and  kinematic misalignments  are often  a
  function  of the  radius. Therefore  it is  generally impossible  to
  characterize the full shape of  a single elliptical galaxy with only
  one or two parameters.

\item Galaxy discs are, in  general, well represented by nearly oblate
  models  with $Q\sim0.9$.  Their  intrinsic flattening  is also  well
  constrained to values spanning $0.2<F<0.3$.

\item  The population  of galaxy  bulges can  be modelled  as slightly
  triaxial ellipsoids with  a tendency to be  oblate.  This population
  has   typical  intrinsic   flattenings  of   $F\sim0.65$.   However,
  individual galaxies can have a variety of intrinsic flattenings with
  some extreme  cases sticking out  the plane  of the disc,  these are
  called polar bulges.

\item The distribution  of the triaxiality parameter  of galaxy bulges
  is  strongly  bimodal. This  bimodality  is  driven by  bulges  with
  S\'ersic index  $n>2$.  According to numerical  simulations they can
  be  explained  assuming a  combination  of  major dissipational  and
  dissipationless mergers during their formation.

\item Despite previous findings showing  a triaxial bulge in the Milky
  Way, more recent studies have found that is more likely a boxy bulge
  produced by the vertical instabilities of the Galactic bar. Owing to
  recent kinematic measurements a classical bulge with mass $>15\%$ of
  the disc mass can be ruled out.

\end{itemize}

Despite the study of the intrinsic  shape of elliptical galaxies has a
long track record, our knowledge of the 3D shape of bulges is still in
its infancy. Therefore,  further work on the topic is  needed to fully
exploit its possibilities.  A few  guidelines to this future prospects
are outlined in the following:

\begin{itemize}
\item From a photometric point of view, even if new methodologies have
  been developed they  need to be applied to larger  samples of galaxy
  bulges.  The  number of  elliptical  galaxies  recently analysed  to
  recover their intrinsic shape is  several orders of magnitude larger
  than the current  samples of galaxy bulges.  Large number statistics
  have led  to the  discovery of  important relations  for ellipticals
  galaxies,  such  as  the  different   shapes  of  bright  and  faint
  ellipticals, and similar  studies can be crucial  for galaxy bulges.
  This  is  particularly relevant  in  the  current picture  of  bulge
  formation with a different  population of classical and pseudobulges
  dependent of the galaxy mass \citep{fisherdrory11}.

\item  An even  more promising  path, already  explored in  elliptical
  galaxies, is  the use of  combined information from  photometric and
  kinematic data.   In particular,  the common  use of  integral field
  spectroscopy is now providing an exquisite detail of the stellar and
  gaseous  kinematics on  large  sample of  galaxies.  This wealth  of
  information  together  with  the  development  of  galaxy  dynamical
  modelling can provide a proper  understanding of the intrinsic shape
  of galaxy bulges.

\item It  is doubtless  that the comparison  of the  derived intrinsic
  shape of bulges with the state-of-the-art numerical simulations is a
  promising way  to gain  insights on the  formation and  evolution of
  bulges. However, there  is still a lack of simulations  with a large
  variety  of   initial  and  physical  conditions   interested  on  a
  structural  analysis  of the  different  galaxy  components, and  in
  particular, in the intrinsic shape evolution of galaxy bulges.

\item  Historically, galaxy  bulges were  thought as  single-component
  objects at  the centre of  galaxies. This picture is  now questioned
  since different bulge types with different formation paths have been
  found coexisting  within the same galaxy  \citep[see][and references
    therein]{mendezabreu14}.  A proper  separation of different bulges
  types, as well as the  identification of possible unresolved nuclear
  structures  such as  bars,  rings,  etc, must  be  accounted for  to
  improve our knowledge on bulge formation and evolution.

\item The study of the intrinsic  shape of elliptical galaxies at high
  redshift has recently suffered a boost thanks to the arrival of high
  spatial resolution  surveys on large fields  of view \citep[see][and
    references therein]{chang13}.  This kind of studies can provide an
  in-situ  view  of galaxy  evolution  and  their application  to  the
  intrinsic shape  of bulges will be  key to further progress  on this
  topic.

\end{itemize}

\begin{acknowledgement}
I would like to thank the  editors E. Laurikainen, R.F.  Peletier, and
D. Gadotti for  their invitation to take part in  this volume. I would
also  like to  thank  A. de  Lorenzo-C\'aceres and  J.   Argyle for  a
careful reading of this manuscript.  JMA acknowledges support from the
European Research Council Starting Grant (SEDmorph; P.I. V. Wild).
\end{acknowledgement}
\newpage
\bibliographystyle{mn2e}
\bibliography{reference}
\clearpage

\end{document}